\documentclass[twocolumn]{aastex6}
\usepackage{amsmath}
\pdfoutput=1

\shorttitle{Rate of Protostellar Outbursts in Orion}
\shortauthors{Fischer, Safron, \& Megeath}

\begin{document}

\newcommand{\Herschel}{\textit{Herschel}}
\newcommand{\Spitzer}{\textit{Spitzer}}
\newcommand{\WISE}{\textit{WISE}}

\title{Constraining the Rate of Protostellar Accretion Outbursts\\ in the Orion Molecular Clouds}

\author{
William J. Fischer,$^1$
Emily Safron,$^2$ and
S. Thomas Megeath$^3$
}
\affiliation{
$^1$Space Telescope Science Institute, Baltimore, MD, USA; wfischer@stsci.edu \\
$^2$Department of Physics and Astronomy, Louisiana State University, Baton Rouge, LA, USA \\
$^3$Ritter Astrophysical Research Center, Department of Physics and Astronomy, University of Toledo, Toledo, OH, USA
}

\begin{abstract}
Outbursts due to dramatic increases in the mass accretion rate are the most extreme type of variability in young stellar objects. We searched for outbursts among 319 protostars in the Orion molecular clouds by comparing 3.6, 4.5, and 24~\micron\ photometry from the \textit{Spitzer Space Telescope} to 3.4, 4.6, and 22~\micron\ photometry from the \textit{Wide-field Infrared Survey Explorer} (\WISE) obtained $\sim6.5$ yr apart. Sources that brightened by more than two standard deviations above the mean variability at all three wavelengths were marked as burst candidates, and they were inspected visually to check for false positives due primarily to the reduced angular resolution of \WISE\ compared to \Spitzer. We recovered the known burst V2775~Ori (HOPS~223) as well as a previously unknown burst, HOPS 383, which we announced in an earlier paper. No other outbursts were found. With observations over 6.5 yr, we estimate an interval of about 1000 yr between bursts with a 90\% confidence interval of 690 to 40,300~yr. The most likely burst interval is shorter than those found in studies of optically revealed young stellar objects, suggesting that outbursts are more frequent in protostars than in pre-main-sequence stars that lack substantial envelopes.
\end{abstract}


\section{INTRODUCTION}\label{s.intro}

\setcounter{footnote}{3}

Young stellar objects (YSOs) display a wide range of photometric variability. Common sources of variability include the rotation of the star, which can carry hot or cool spots across the line of sight \citep{her94}, the rotation of the protoplanetary disk, which can carry orbiting disk structures across the line of sight \citep{bou07,mor11}, changes in the structure of the disk \citep{ric12,wol13,fla12,fla13}, the influence of a companion star \citep{muz13}, or changes in the foreground extinction \citep{chi15}. The most extreme examples of variability, however, are due to changes in the luminosity of the system. While the luminosity of a low-mass pre-main-sequence star is expected to change gradually over $\sim10^6$~yr as it contracts toward the main sequence, the luminosity generated by disk material accreting onto the star may vary on much shorter timescales due to changes in the accretion rate.

Episodic accretion is a phenomenon in which the accretion rate of a forming star rapidly increases by as much as several orders of magnitude \citep{har96}. In the case of protostars, where accretion can dominate the total luminosity, it is one potential solution to the classic luminosity problem, where luminosities predicted for protostars that accrete at a constant rate throughout the formation period are greater than those typically observed \citep{ken90,eva09}. With episodic accretion, the protostar is usually less luminous than such predictions indicate, in agreement with observations, but it spends enough time at an elevated luminosity for its luminosity averaged over the entire period to agree with predictions. Many examples of episodic accretion have been observed since the first discovery of an outburst, that of FU Ori, in 1936 \citep{wac39}. While outbursts clearly occur, the fraction of a typical star's main-sequence mass that is accumulated during bursts is still poorly constrained.

Investigating the overall importance of bursts to star formation, \citet{dun10} showed that models featuring episodic accretion were successful in reproducing the bolometric luminosity and temperature (BLT) distribution, a common evolutionary diagram, of 1024 nearby YSOs. \citet{off11}, on the other hand, found that models with a limited role for episodic accretion but with constant accretion times reproduced observations better than those with constant accretion rates. To explain the BLT distribution of 315 Orion protostars, \citet{fis17b} invoked exponentially declining accretion rates over the star formation period with, again, a limited role for episodic accretion. 

The reviews of \citet{rei10} and \citet{aud14} tabulate the known outbursts, discuss their observational characteristics, and explore potential triggering mechanisms. The two main types of bursts are EX Lupi and FU Orionis events. EX Lupi bursts are relatively small in amplitude and short-lived, but they are also known to recur in the same star \citep{asp10}. 

Conversely, the amplitudes of FU Orionis bursts are greater. Their decay timescales are far longer, more than several decades, but with significant variation in their post-burst light curves. Bursts of this type may recur in the same star, but the timescales are too long for this to have yet been observed. Several outbursts do not fit cleanly into either class, such as V2492 Cyg \citep{cov11} or V1647 Ori \citep{muz05,fed07,con17a}, indicating that the usual division of bursts into two classes is a simplification.

The predominant explanation for the jump in accretion rate invokes disk instabilities, either thermal or gravitational, triggered by growth in the disk mass due to accumulation of infalling envelope material. In thermal instability models, the gas accumulates in the disk as it is blocked from accreting onto the star by an orbiting planet or a dead zone, eventually leading to a rapid rise in temperature and accretion rate as the material falls onto the star \citep{lod04,arm01,zhu10}. In the gravitational instability models, the growing disk becomes unstable and fragments into clumps, which migrate into the inner disk. The system begins a luminosity outburst when the clumps accrete onto the star \citep{vor05,vor15}. 

In this work we use the common classification system for YSOs. Classes I, II, and III were defined by \citet{lad87}, and the system was extended with the Class 0 \citep{and93} and flat-spectrum \citep{gre94} categories. For this study, we consider Class~0, Class~I, and flat-spectrum YSOs to be protostars, because their circumstellar envelopes make an important contribution to their observed properties and subsequent evolution. \citet{fur16} provide detailed criteria for distinguishing among these classes and compare the properties of the different classes of protostars in Orion.

Though more widely studied within the last two decades, many characteristics of accretion bursts are still not well constrained. This is due in part to the serendipitous nature of the discovery of most outbursts, which limits the effectiveness of statistical techniques in estimating their frequency. Further, detections of protostellar outbursts are rare compared to detections of other types of YSO variability. Observational studies of large samples of YSOs are essential in constraining the characteristics of the outburst process and in guiding theoretical work.

The advent of wide-area, IR surveys of large numbers of diverse star-forming regions makes systematic searches for accretion bursts feasible. \citet{sch13}, for example, used two epochs of photometry for about 8000 YSOs covering a wide range of masses and ages, searching for luminosity increases that satisfy specific outburst criteria, with the ultimate goal of constraining the frequency of outbursts in their sample. They found a burst interval of 5000 to 50,000~yr. For all sample sizes and time baselines studied to date, however, the upper bound of the burst interval is poorly constrained \citep{hil15}.

Here we present the results from an analysis of 319 well characterized protostars in the Orion molecular cloud complex, using two epochs of mid-IR photometry from the {\it Spitzer Space Telescope} \citep{wer04} and the {\it Wide-field Infrared Survey Explorer} (\WISE; \citealt{wri10}). This survey is the largest to date for protostellar outbursts, given the large population of young YSOs in the Orion region. Like \citet{sch13}, we will use the results of this survey to estimate the frequency of outbursts between the two epochs.

\section{DATA AND SAMPLE SELECTION}\label{s.data}

As part of the \Spitzer\ Orion Survey \citep{meg12,meg16}, fields in the Orion A and B molecular clouds were mapped in 2004 and 2005 with the Infrared Array Camera (IRAC; \citealt{faz04}) and the Multiband Imaging Photometer for \Spitzer\ (MIPS; \citealt{rie04}). Images and photometry were obtained at five wavelengths, including bands centered at 3.6 (IRAC~1), 4.5 (IRAC~2), and 24~\micron\ (MIPS~1). The FWHM of the point-spread function at each wavelength is approximately 2\arcsec\ for the IRAC channels and 6\arcsec\ for MIPS. \citet{meg12} show that the $5\sigma$ detection limits in each band are typically 17 mag in IRAC~1 and IRAC~2 and 9 mag in MIPS~1; these limits vary across the Orion survey due to confusion with spatially varying nebulosity. They also show the coverage of the maps and the dates on which they were obtained.

The \Spitzer\ data were used to identify targets for the \Herschel\ Orion Protostar Survey (HOPS; \citealt{man13,stu13,fur16,fis17b}), an open-time key program of the {\it Herschel Space Observatory} to obtain 70 and 160~\micron\ imaging and 50--200~\micron\ spectra of Orion protostars. \citet{fur16} presented 1 to 870~\micron\ SEDs and model fits for 330 HOPS targets in Orion. These are protostellar candidates that were imaged by HOPS and detected in the \Herschel\ 70~\micron\ images. Of the 330 YSOs, \citet{fur16} concluded from SED analysis that 319 are protostars and the remaining 11 are Class II YSOs. These 319 protostars, those that appear in Table 1 of \citet{fur16} and have Class designations of 0, I, or flat, constitute the sample for this work. Their coordinates and SED properties are tabulated there as well.

\begin{deluxetable}{ccccc}
\tablewidth{\hsize}
\tablecaption{Two-Epoch MIPS 24 \micron\ Photometry\label{t.twophot}}
\tablehead{\colhead{} & \multicolumn{2}{c}{2004--2005\tablenotemark{a}} & \multicolumn{2}{c}{2008\tablenotemark{b}} \\ \colhead{HOPS ID} & \colhead{Mag.} & \colhead{Unc.\tablenotemark{c}} & \colhead{Mag.} & \colhead{Unc.\tablenotemark{c}}}
\startdata
86 & 1.788 & 0.055 & 2.105 & 0.057 \\
87 & 2.490 & 0.055 & 2.396 & 0.061 \\
88 & 3.268 & 0.055 & 2.840 & 0.066 \\
89 & 3.779 & 0.055 & 3.645 & 0.064 \\
90 & 1.842 & 0.055 & 1.911 & 0.072 \\
91 & 4.095 & 0.056 & 3.821 & 0.069 \\
92 & 0.451 & 0.055 & 0.571 & 0.057 \\
93 & 4.311 & 0.055 & 4.476 & 0.066 \\
94 & 1.332 & 0.055 & 1.391 & 0.077 \\
95 & 5.167 & 0.056 & 5.174 & 0.062 \\
96 & 4.252 & 0.055 & 4.222 & 0.056 \\
99 & 5.256 & 0.057 & 5.459 & 0.077 \\
100 & 5.977 & 0.057 & 5.779 & 0.062 \\
102 & 3.700 & 0.056 & 3.706 & 0.058 \\
105 & 5.349 & 0.057 & 5.442 & 0.066 \\
107 & 1.517 & 0.056 & 1.526 & 0.058 \\
204 & 3.388 & 0.061 & 3.507 & 0.055 \\
206 & 4.087 & 0.055 & 3.746 & 0.055 \\
207 & 3.810 & 0.055 & 3.582 & 0.055 \\
208 & 6.787 & 0.059 & 7.082 & 0.061 \\
209 & 4.473 & 0.055 & 4.481 & 0.055 \\
210 & 2.854 & 0.055 & 2.901 & 0.055 \\
211 & 5.996 & 0.057 & 6.071 & 0.057 \\
213 & 2.622 & 0.056 & 2.729 & 0.057 \\
214 & 5.030 & 0.055 & 5.180 & 0.056 \\
215 & 3.545 & 0.055 & 3.563 & 0.055 \\
216 & 3.147 & 0.055 & 3.231 & 0.056 \\
219 & 2.546 & 0.056 & 2.505 & 0.056 \\
220 & 4.755 & 0.057 & 4.463 & 0.056 \\
221 & 0.981 & 0.055 & 1.022 & 0.055 \\
223 & 2.548 & 0.055 & 0.389 & 0.056 \\
224 & 4.580 & 0.056 & 4.423 & 0.056 \\
225 & 3.597 & 0.059 & 3.444 & 0.055 \\
226 & 3.752 & 0.059 & 3.514 & 0.055 \\
227 & 4.378 & 0.055 & 4.445 & 0.055 \\
228 & 0.410 & 0.055 & 0.714 & 0.055 \\
232 & 2.536 & 0.056 & 2.290 & 0.061 \\
235 & 1.429 & 0.056 & 1.580 & 0.067 \\
383 & 7.875 & 0.057 & 4.015 & 0.057 \\
\enddata
\tablenotetext{a}{Used in this paper.}
\tablenotetext{b}{Used in \citet{fur16}.}
\tablenotetext{c}{All uncertainties include a 5\% floor.}
\end{deluxetable}

Our \Spitzer\ photometry is from the 2004--2005 campaigns discussed above. The data are published in Table~2 of \citet{fur16}, except that those authors used more recent MIPS photometry for 39 of the protostars and more recent IRAC photometry for HOPS~383. In Table~\ref{t.twophot}, we list the original and previously unpublished 2004--2005 MIPS photometry used here for these sources as well as their previously published 2008 MIPS photometry. Both epochs of IRAC photometry for HOPS 383 appear in \citet{saf15}.

The \WISE\ mission surveyed the entire sky in four bands, including 3.4 (W1), 4.6 (W2), and 22~\micron~(W4). \WISE\ mapped Orion at all wavelengths in 2010 March, and after the exhaustion of cryogen, the region was mapped again at W1 and W2 in 2010 September. This work uses the AllWISE version of the \WISE\ point-source catalog, which incorporates both epochs. The FWHM of the point-spread function for W1 and W2 is approximately 6\arcsec, while that for W4 is approximately 12\arcsec. The quoted $5\sigma$ sensitivities for AllWISE are 16.9 mag in W1, 15.9 mag in W2, and 8.0 mag in W4, although these sensitivities can be considerably worse in a high-background region such as Orion.\footnote{See the discussion at \url{http://wise2.ipac.caltech.edu/docs/release/allwise/expsup/sec2\_3a.html}.} The three \WISE\ bands of interest have response functions similar to \Spitzer's IRAC~1, IRAC~2, and MIPS~1 bands, respectively. We find at most a weak correlation between the intrinsic \Spitzer\ color of a source and the difference between its \Spitzer\ and \WISE\ magnitudes, so we conclude that differences in the \Spitzer\ and \WISE\ response functions are a minor effect compared to true outbursts, and we directly compare the two sets of magnitudes. For simplicity we use 3.6, 4.5, and 24~\micron\ to refer to the central wavelengths for both telescopes.

Comparison of the two data sets allows the discovery of variability on a timescale of 5 to 6 yr. Due to the larger point-spread functions of \WISE, blending with nearby sources is a concern, and visual inspection is needed to conclusively identify variable sources.

Our process for matching \WISE\ point sources to the HOPS protostars is that described in \citet{fis16a}. We used the NASA/IPAC Infrared Science Archive\footnote{See \url{http://irsa.ipac.caltech.edu}.} to search AllWISE for the closest match to each \Spitzer\ position. For 227 of 319 protostars, there was a match within 1\arcsec; these were automatically paired with their \Spitzer\ counterparts. For 32 protostars, there was no \WISE\ source within 10\arcsec; these were automatically marked as having no counterpart. For the 60 sources at intermediate separations, we manually inspected the images and found that the larger offset was due either to scattered light from the \Spitzer\ point source or to a blend in \WISE\ of two or more distinct \Spitzer\ sources. We retained the 44 matches where the \WISE\ flux was judged to be mainly due to the \Spitzer\ source in question (all with separations less than 7\arcsec) and rejected the other 16. Therefore, 271 of the 319 protostars have \WISE\ counterparts.

\begin{deluxetable}{lccccc}
\tablecaption{HOPS Protostars with \WISE\ Counterparts\label{t.detect}}
\tablewidth{\hsize}
\setlength{\tabcolsep}{2.4pt}
\tablehead{\colhead{} & \colhead{3.6 \micron} & \colhead{4.5 \micron} & \colhead{24 \micron} & \colhead{All}}
\startdata
\Spitzer\ and \WISE\ & 248 & 260 & 249 & 233 \\
\Spitzer\ only & 3 & 0 & 20 & \nodata \\
\WISE\ only & 16 & 10 & 0 & \nodata \\
No detection & 4 & 1 & 2 & \nodata \\
\tableline
Mean difference (mag)\tablenotemark{a} & $-0.283$ & $0.356$ & $0.013$ \\
Std.\ dev.\ of difference (mag)\tablenotemark{a} & $\phm{-}0.646$ & $0.590$ & $0.620$ \\
\enddata
\tablenotetext{a}{Mean and standard deviation of \Spitzer\ magnitude minus\\ \WISE\ magnitude for sources detected in both catalogs.}
\end{deluxetable}

A comparison of the detections by \Spitzer\ and \WISE\ is shown in Table~\ref{t.detect}. Of the 271 protostars with \WISE\ counterparts, 233 are detected by both \Spitzer\ and \WISE\ in all three bands of interest. Considering the bands independently, in each band 91\% to 96\% of the 271 protostars are detected in both catalogs. Of those that were detected in both, the mean difference in magnitude is slightly negative at 3.6~\micron, slightly positive at 4.5~\micron, and nearly zero at 24~\micron. The standard deviations in these differences are about 0.6~mag in all bands. Throughout this work, we subtract \WISE\ magnitudes from \Spitzer\ magnitudes so that positive numbers correspond to an increase in brightness from one epoch to the next.

To identify outburst candidates, we use these typical differences as a guide and flag protostars that became substantially brighter between the \Spitzer\ and \WISE\ maps or were detected in the later \WISE\ map after not being detected by \Spitzer. We then manually inspect these candidates to assess whether the different angular resolutions of the two telescopes play a role in the reported changes in brightness. In the following section we present and justify outburst criteria.

\section{CANDIDATE CRITERIA}\label{s.crit}

Depending on the underlying physics, YSO variability is characterized by a range of amplitudes and color changes. Here we are interested in major increases in the accretion rate from the disk onto the star. In this section we discuss the expected effect of such increases on the SED at 3.6, 4.5, and 24~\micron\ and develop criteria for identifying outburst candidates. We argue that a large fractional change in the accretion rate will induce fractional changes of a similar magnitude in the flux densities at these wavelengths.

The accretion rate from the disk onto the star $\dot{M}$ is proportional to the accretion luminosity $L_{\rm acc}$, where $L_{\rm acc} = \eta\, G M_* \dot{M}/R_*$. Here $G$ is the gravitational constant, $M_*$ and $R_*$ are the mass and radius of the central object, and $\eta$ is a constant of order unity that depends on the details of the accretion process. A large increase in $\dot{M}$ will drive a proportional change in the accretion luminosity, since the stellar parameters do not change on timescales short enough to be relevant and changes in $\eta$ are expected to be small ($\lesssim20\%$) compared to the change in $\dot{M}$, which is typically one order of magnitude or more \citep{bar12}.

The total luminosity of the system $L$ is the sum of the accretion luminosity and the luminosity of the central star $L_*$. Like other parameters intrinsic to the star, $L_*$ is not expected to change appreciably due to the outburst. The ratio of the total luminosity in outburst to that in the quiescent state is then $(L_{{\rm acc}, b} + L_*)/(L_{{\rm acc}, q} + L_*)$. This is a lower limit to the ratio of accretion luminosities, $L_{{\rm acc}, b}/L_{{\rm acc}, q}$, and it approaches the latter ratio when both accretion luminosities are much larger than that of the star, the expectation early in protostellar evolution \citep{ada87,and93,fis17b}. Therefore, a large fractional increase in the total luminosity of the system implies even larger fractional increases in the accretion luminosity and accretion rate.

In the SED, a change in the total luminosity may affect the flux density differently at different wavelengths. Figure 11 of \citet{fur16} shows how, as the luminosity of a protostar increases from 0.1 to 303 $L_\sun$ in steps of about half an order of magnitude, the fractional change in flux density predicted by a radiative transfer model at 3.6, 4.5, and 24 \micron\ is similar to the fractional change in luminosity. This was also indicated by \citet{ken93}, who calculated that the peak wavelength of a protostellar SED varies as $L^{-1/12}$; i.e., a change in luminosity primarily affects the overall flux in the SED, not its shape. Finally, \citet{sch13} analyzed model SEDs for prototypical Class I and II sources at various accretion rates. They found that, between 2 and 5 \micron, the flux increases by a factor of 10 or more when the accretion rate increases from zero to $10^{-6}~M_\sun~{\rm yr}^{-1}$ or more. These lines of reasoning suggest that dramatic increases in luminosity can be detected by searching for simultaneous large increases in flux density at all three wavelengths of interest.

The typical short-term near- or mid-IR variations in large samples of YSOs, which may be due to rotational modulation by spots on the star or inner disk inhomogeneities, are in the range of 0.1--0.6 mag \citep{mor11,fla12,meg12} and can be strongly wavelength dependent. The slightly different \Spitzer\ and \WISE\ bandpasses can mimic variability between the two surveys of up to a few tenths of a magnitude. Low-level variability and differences in the bandpasses are likely responsible for the distributions in magnitude differences presented in Table~\ref{t.detect}, where the mean and standard deviation in each band are of order a few tenths of a magnitude. 

To filter out low-level variability and differences due to the different bandpasses, we search for protostars with brightenings greater than two standard deviations above the mean at all three wavelengths of comparison, or 1.01~mag at 3.6~\micron, 1.54~mag at 4.5~\micron, and 1.25~mag at 24~\micron. We also consider protostars that exceed the threshold at one or two wavelengths while changing from a non-detection to a detection at the other(s). These changes correspond to an increase in luminosity of a factor of approximately 2.5--4. Since the variations in the photometry are dominated by low-level variability as well as by potential source confusion in the \WISE\ data, the two-sigma threshold should not be considered the statistical significance of the bursts. Instead, it gives criteria for distinguishing large bursts from smaller-scale variability and systematic effects in the data. The thresholds are expected to pick up moderate outbursts as well as those in which neither epoch of photometry exactly catches the minimum or maximum brightness.

\section{ANALYSIS OF BRIGHTNESS CHANGES}\label{s.analysis}

In this section we plot the difference in magnitude of each protostar at the three wavelengths of comparison, identify those that brightened by more than the threshold at one or more wavelengths, and examine false positives. 

\begin{figure}
\includegraphics[width=\hsize]{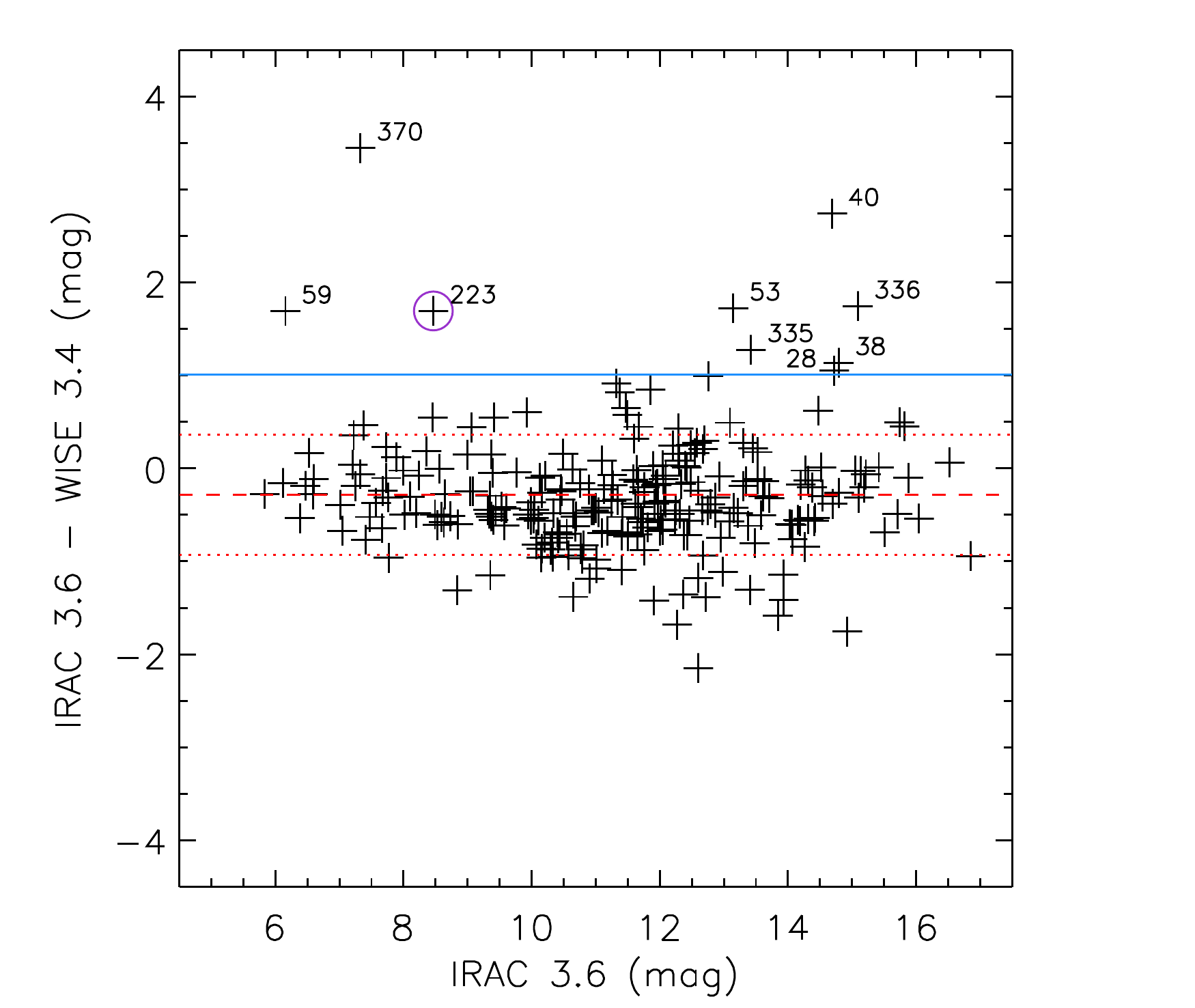}
\caption{Magnitude differences between IRAC~1 and W1 plotted against IRAC~1 magnitudes for the 248 protostars detected in both bands. The dashed red line is the mean difference, and dotted red lines show the mean difference plus or minus one standard deviation. The solid blue line shows the mean plus two standard deviations (1.01~mag), the outburst threshold. Labeled protostars exceeded the threshold at this wavelength; the encircled protostar (HOPS 223) is the only one detected in both bands that exceeds the thresholds at all three wavelengths.\label{f.diffmag1}}
\end{figure}

Figure~\ref{f.diffmag1} plots the IRAC~1 $-$ W1 magnitude against the IRAC~1 magnitude for the 248 protostars detected at both epochs. Nine protostars lie above the threshold for outbursts; these points are labeled with their HOPS numbers. Five of them satisfy the requirement only at this wavelength; HOPS 28, 335, and 370 satisfy the requirement at this and one other wavelength; and HOPS~223 satisfies the requirement at all three wavelengths. 

\begin{figure}
\includegraphics[width=\hsize]{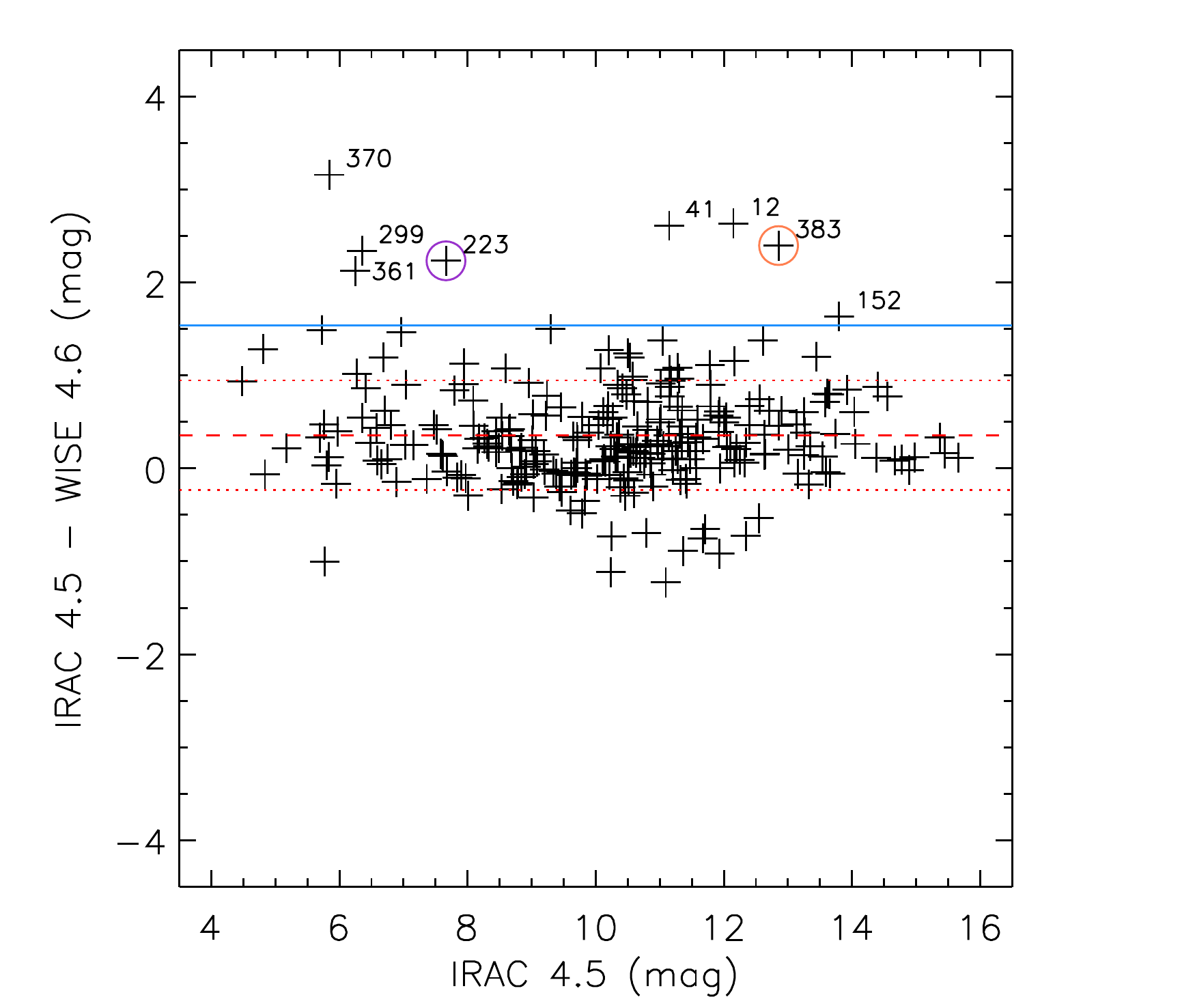}
\caption{Magnitude differences between IRAC~2 and W2 plotted against IRAC~2 magnitudes for the 260 protostars detected in both bands. Lines, annotations, and the purple circle have the same meaning as in Figure~\ref{f.diffmag1}, where the outburst threshold at this wavelength is 1.54~mag. The object encircled in orange is HOPS 383, which exceeds the threshold at two of three wavelengths and was not detected in the 3.6 \micron\ \Spitzer\ band.\label{f.diffmag2}}
\end{figure}

Figure~\ref{f.diffmag2} plots the IRAC~2 $-$ W2 magnitude against the IRAC~2 magnitude for the 260 protostars detected in both bands. Eight protostars lie above the threshold for outbursts. Five of them satisfy the requirement only at this wavelength, and two of them satisfy the requirement at this and one other wavelength. Of these two, HOPS 370 brightened at 3.6 and 4.5 \micron\ but not 24~\micron. The other, HOPS 383, was not detected at 3.6~\micron\ by \Spitzer\ but brightened at the other two wavelengths. Again, HOPS 223 satisfies the requirement at all three wavelengths. 

\begin{figure}
\includegraphics[width=\hsize]{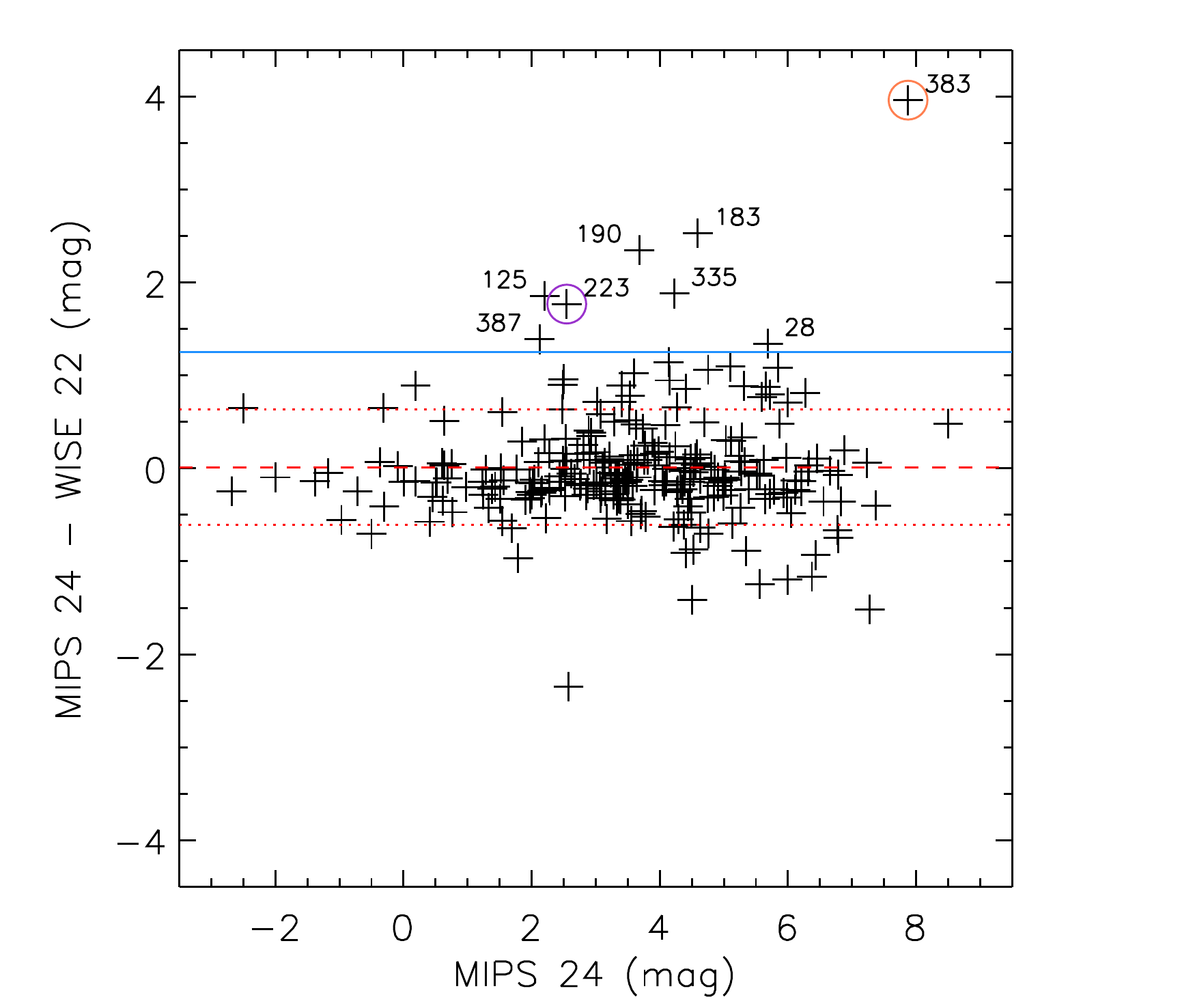}
\caption{Magnitude differences between MIPS~1 and W4 plotted against MIPS~1 magnitudes for the 249 protostars detected in both bands. Lines, annotations, and circles have the same meaning as in Figure~\ref{f.diffmag2}, where the outburst threshold at this wavelength is 1.25~mag.\label{f.diffmag4}}
\end{figure}

Figure~\ref{f.diffmag4} plots the MIPS~1 $-$ W4 magnitude against the MIPS~1 magnitude for the 249 protostars detected in both bands. Eight protostars lie above the threshold for outbursts. Four of them brightened only at this wavelength, and three of them brightened at one other wavelength. These are HOPS 28, 335, and 383, which are all mentioned above. Again, HOPS 223 satisfies the requirement at all three wavelengths.

\begin{figure}
\includegraphics[width=\hsize]{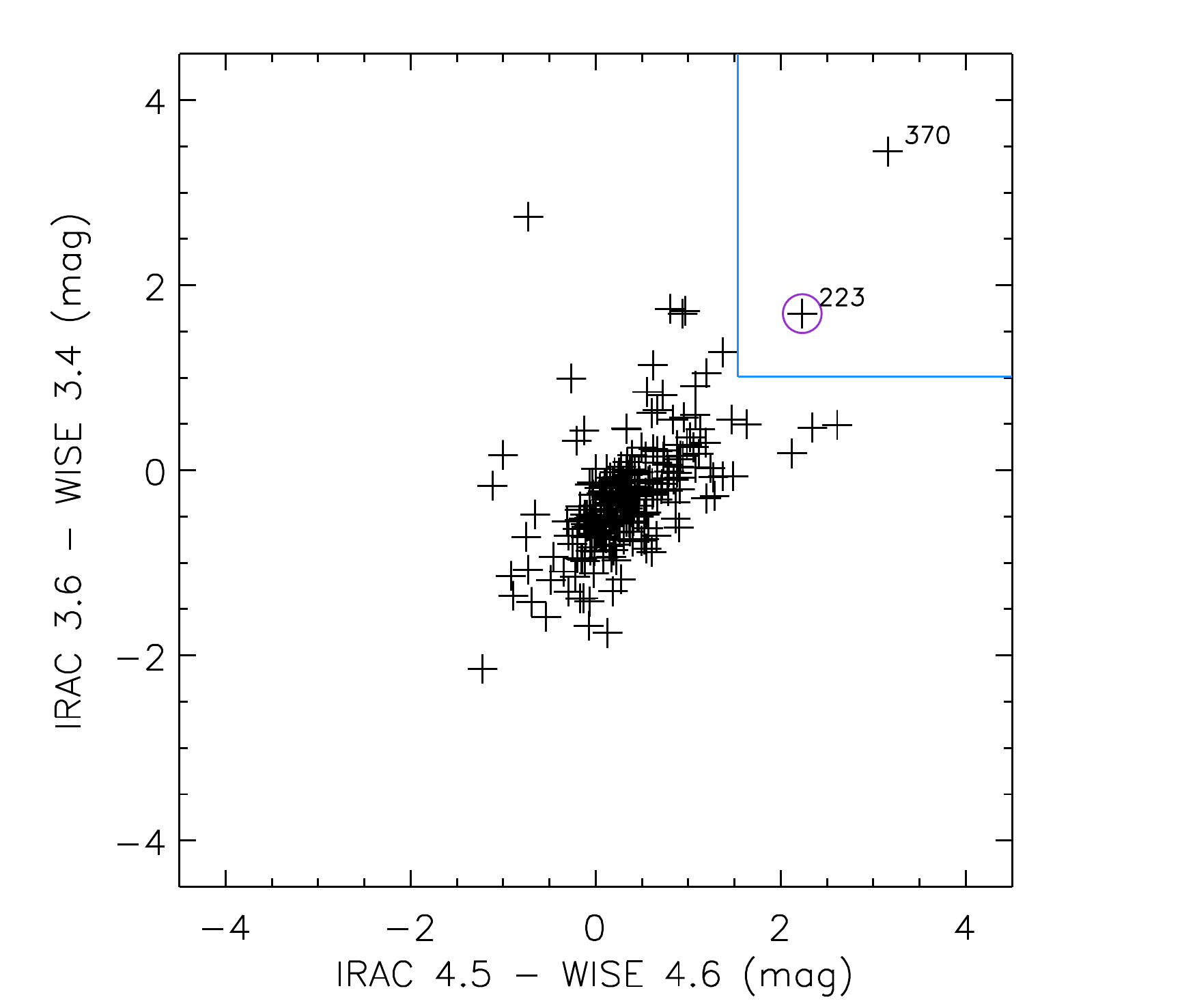}
\caption{Magnitude differences between IRAC 1 and W1 plotted against the magnitude differences between IRAC 2 and W2 for the 248 protostars detected in all four bands. Labeled sources, above and to the right of the blue lines, satisfy the outburst thresholds for both wavelengths. The encircled protostar, HOPS 223, satisfies the outburst thresholds at all three wavelengths.\label{f.diff1diff2}}
\end{figure}

\begin{figure}
\includegraphics[width=\hsize]{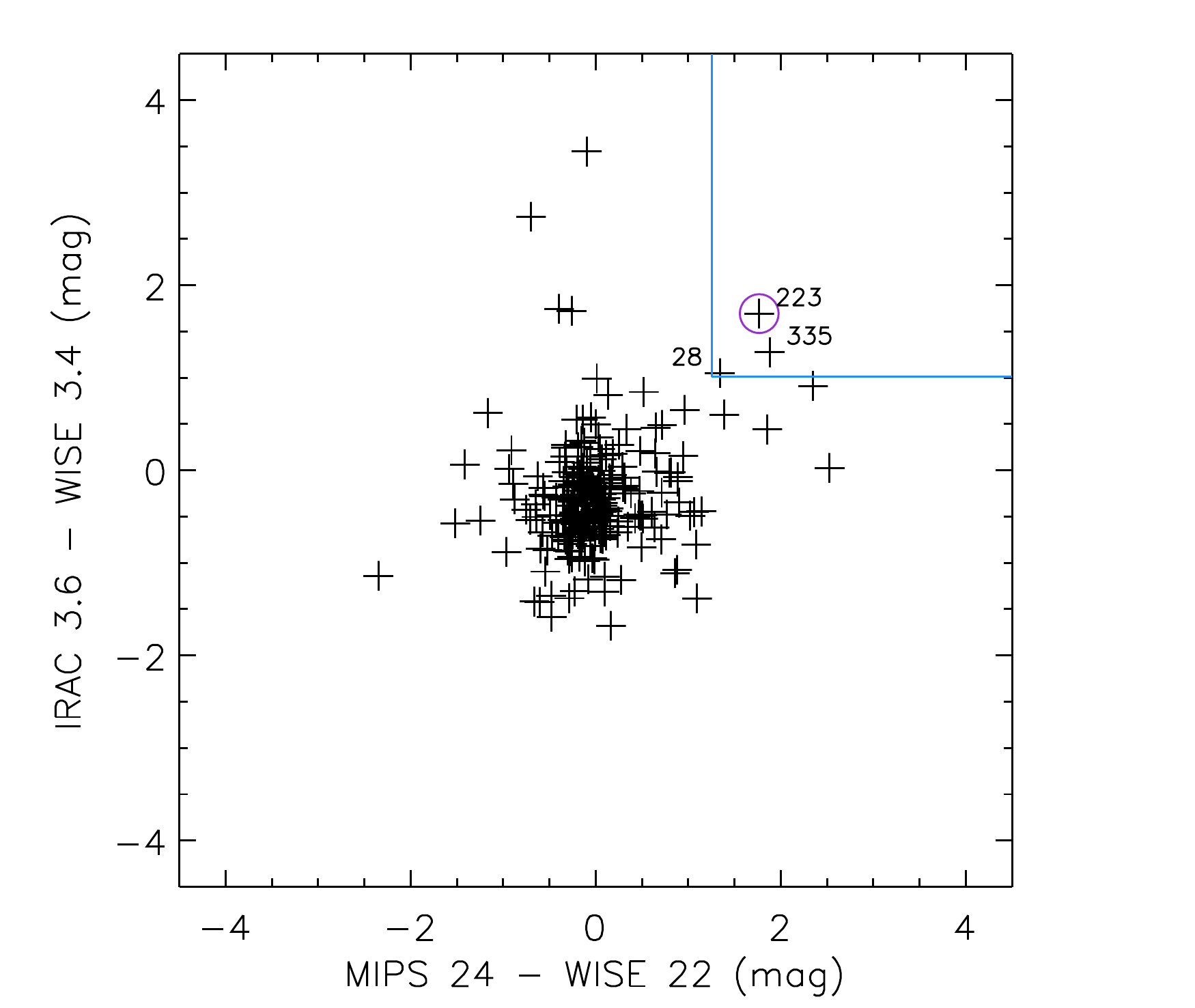}
\caption{Magnitude differences between IRAC 1 and W1 plotted against the magnitude differences between MIPS 1 and W4 for the 233 protostars detected in all four bands. Lines, annotations, and the purple circle have the same meaning as in Figure~\ref{f.diff1diff2}.\label{f.diff1diff4}}
\end{figure}

\begin{figure}
\includegraphics[width=\hsize]{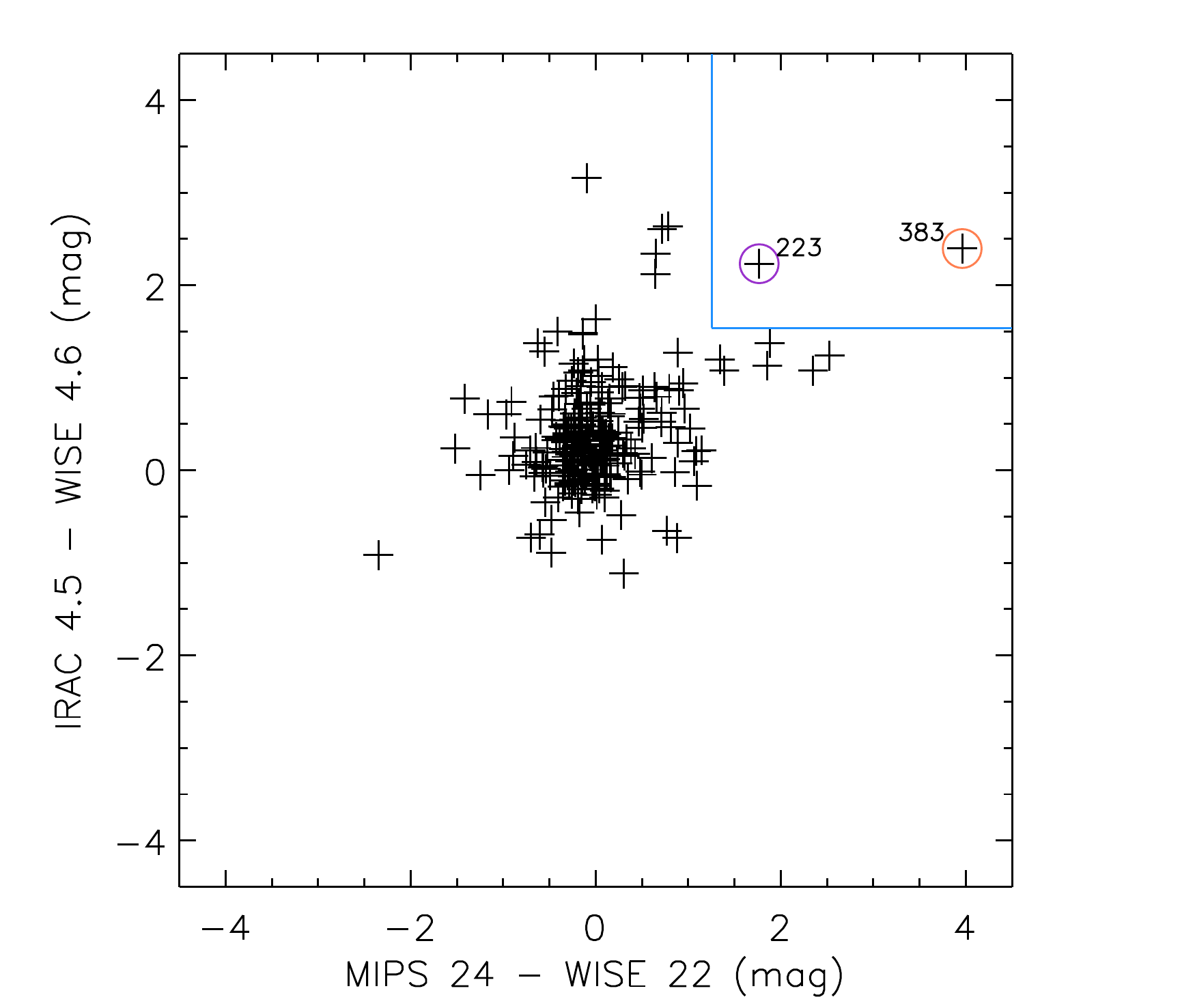}
\caption{Magnitude differences between IRAC 2 and W2 plotted against the magnitude differences between MIPS 1 and W4 for the 241 protostars detected in all four bands. Lines, annotations, and the purple circle have the same meaning as in Figure~\ref{f.diff1diff2}, while the orange circle marks the location of HOPS 383, which satisfies the outburst thresholds at these wavelengths but changed from a non-detection to a detection at 3.6 \micron.\label{f.diff2diff4}}
\end{figure}

Differences at one wavelength are plotted against those at a second wavelength in the next three figures. Figure~\ref{f.diff1diff2} plots the difference in magnitude at 3.6 \micron\ against the difference at 4.5~\micron\ for the 248 protostars detected by \Spitzer\ and \WISE\ at both wavelengths. These differences are moderately well correlated, with a correlation coefficient of 0.60. The wavelengths are close enough that sources are unlikely to brighten or fade substantially at one but not the other. The most flagrant exception, with a putative brightening of more than 2 mag at 3.6~\micron\ but a fading at 4.5 \micron, is HOPS~40, above and slightly to the left of the main cloud of points. It is faint in IRAC~1 (14.7 mag) and contaminated by background emission in W1.

Figure~\ref{f.diff1diff4} plots the difference in magnitude at 3.6 \micron\ against the difference at 24 \micron\ for the 233 protostars detected by \Spitzer\ and \WISE\ at both wavelengths, and Figure~\ref{f.diff2diff4} plots the difference in magnitude at 4.5 \micron\ against the difference at 24 \micron\ for the 241 protostars detected by both telescopes at both wavelengths. These differences are not well correlated, with correlation coefficients of 0.21 and 0.37. The \Spitzer\ 24 \micron\ and \WISE\ 22 \micron\ bands are far enough in wavelength space from the 3--5 \micron\ bands, and the \WISE\ 22 \micron\ band is sufficiently affected by confusion due to its lower angular resolution, that changes in the former do not necessarily track those in the latter.

\subsection{False Positives}

Here we look at the three protostars that were detected at all three wavelengths by both telescopes and satisfy the outburst criteria at just two of the three wavelengths. These are the protostars that lie above and to the right of the blue lines, but are not circled, in the figures that plot one magnitude difference against another. Figure~\ref{f.diff1diff2} shows that HOPS 370 is part of this collection, and Figure~\ref{f.diff1diff4} highlights HOPS 28 and 335. With visual inspection, their putative outbursts can all be rejected as false positives due to contamination of the relatively large \WISE\ point-spread function by nebulosity or nearby point sources. The remaining sources that satisfy only one threshold are also all false positives. They will not be examined in detail here, but we found that they suffer from the same issues described below.

\begin{figure}
\includegraphics[width=\hsize]{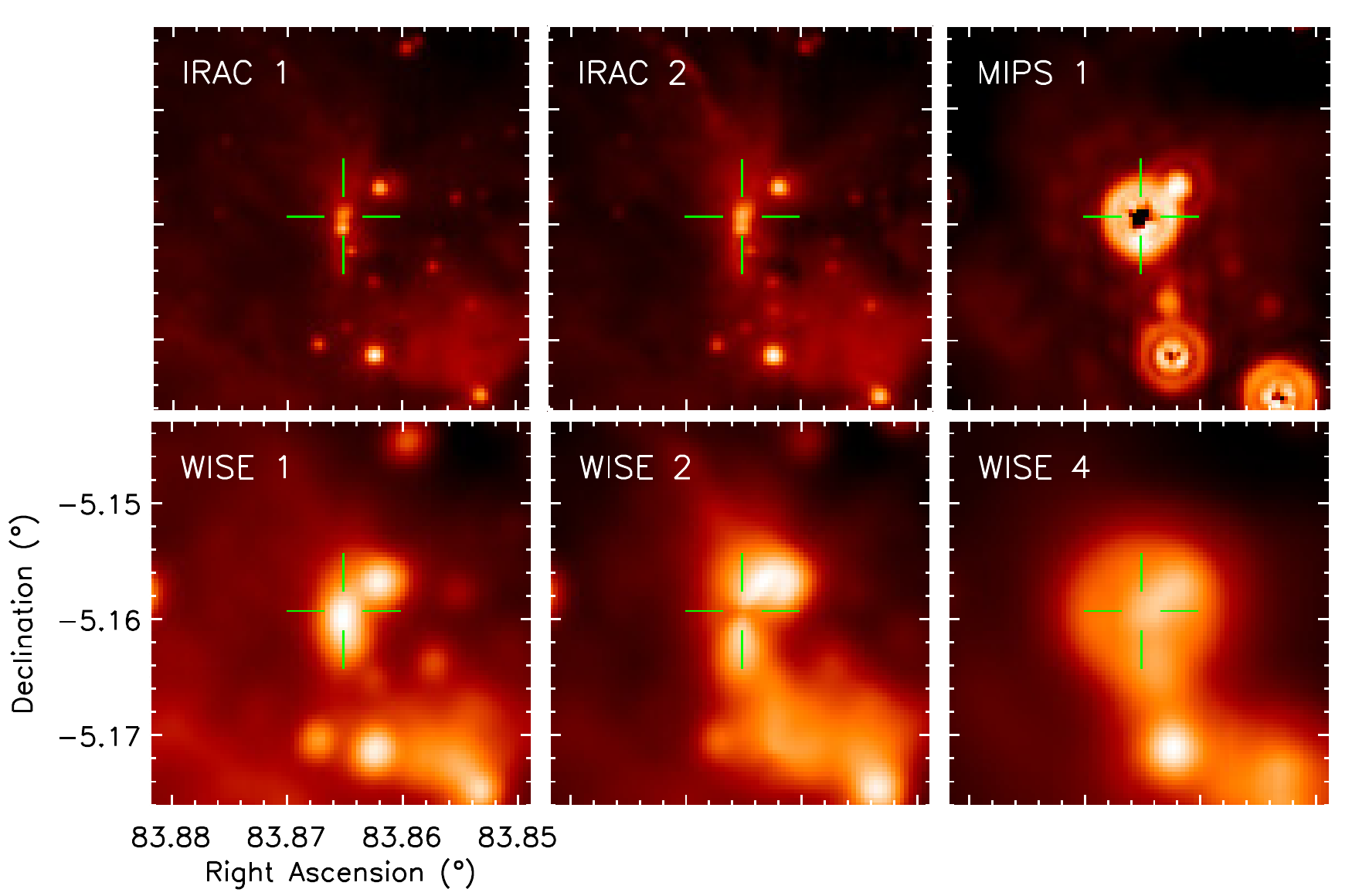}
\caption{HOPS 370 (at the center of the crosshairs) as seen in {\it Spitzer}'s IRAC~1, IRAC~2, and MIPS~1 bands (top row, from left), as well as \WISE's W1, W2, and W4 bands (bottom row, from left).\label{f.370}}
\end{figure}

Figure~\ref{f.370} shows thumbnails of HOPS 370, which satisfies the criteria only at 3.6 and 4.5 \micron. At those wavelengths, it became brighter by 3.9 and 2.7 mag, respectively, while at 24 \micron, it apparently faded by 0.1 mag. HOPS 370 is a binary protostar \citep{nie03} that is barely resolved by IRAC. The saturation evident in the MIPS image does not affect the analysis, since the photometry was obtained by fitting a point-spread function with the saturated pixels masked \citep{kry12}. In W1 the components are blended, while in W2, the sources are visibly separated. In both, extended emission and the bright source just to the northwest (HOPS 66) contribute to the photometry, while in W4, the source is so bright that contamination by these factors is minimal. The erroneously elevated photometry in W1 and W2 make this candidate a false positive.

\begin{figure}
\includegraphics[width=\hsize]{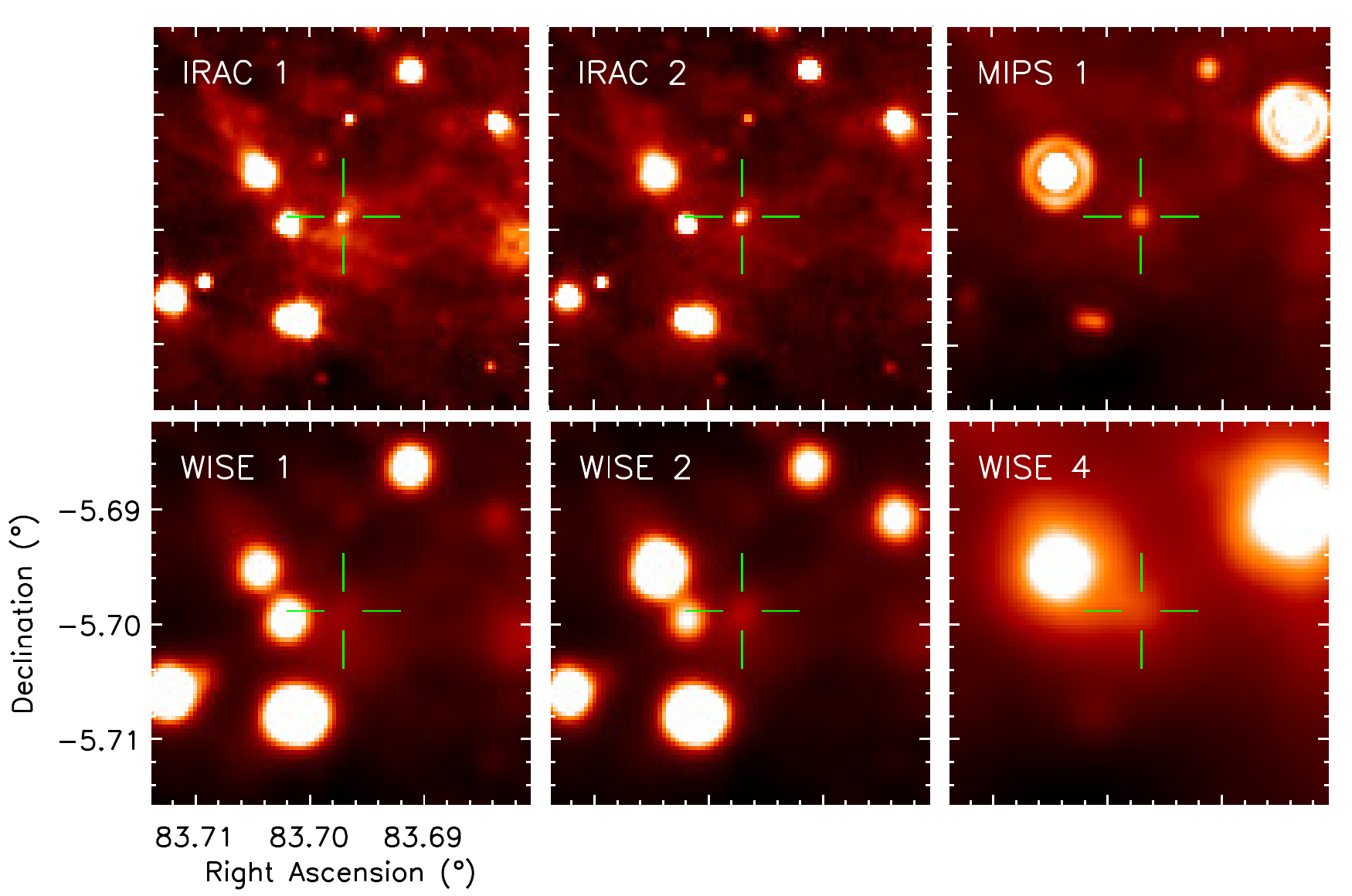}
\caption{HOPS 28 (at the center of the crosshairs) as seen in {\it Spitzer}'s IRAC~1, IRAC~2, and MIPS~1 bands (top row, from left), as well as \WISE's W1, W2, and W4 bands (bottom row, from left).\label{f.28}}
\end{figure}

\begin{figure}
\includegraphics[width=\hsize]{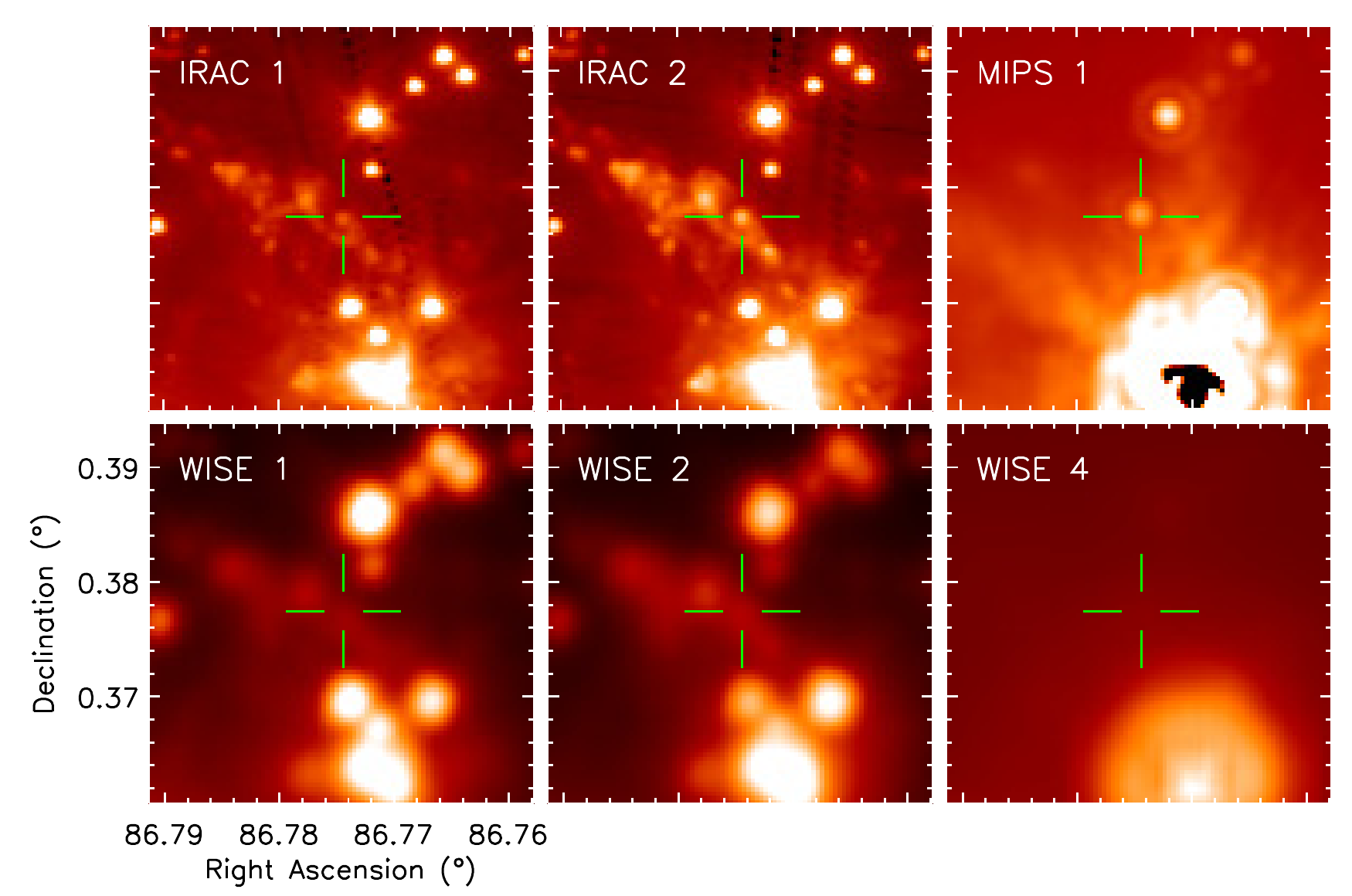}
\caption{HOPS 335 (at the center of the crosshairs) as seen in {\it Spitzer}'s IRAC~1, IRAC~2, and MIPS~1 bands (top row, from left), as well as \WISE's W1, W2, and W4 bands (bottom row, from left).\label{f.335}}
\end{figure}

Figure~\ref{f.28} and Figure~\ref{f.335} show thumbnails of HOPS 28 and HOPS~335, the two protostars that satisfy the criteria only at 3.6 and 24 \micron. HOPS 28 apparently brightened by 1.1 mag at 3.6 \micron, 1.2 mag at 4.5 \micron, and 1.3~mag at 24 \micron. These represent increases of 2.1, 1.4, and 2.1 standard deviations; i.e., it slightly exceeded the criteria at two bands and fell short at the other. HOPS 335 apparently brightened by 1.3 mag at 3.6 \micron, 1.4~mag at 4.5 \micron, and 1.9 mag at 24~\micron, for increases of 2.4, 1.7, and 3.0 standard deviations, coming close to meeting all criteria. In both of these cases, extended emission appears to be contaminating the photometry in W1 and W2, while a nearby, much brighter source contaminates the photometry in W4.

In summary, the larger \WISE\ point spread functions tend to attribute environmental nebulosity to point sources. Blending with nearby point sources is also a concern, particularly at 22 \micron. Visual inspection is a crucial step to identify false outbursts in the comparison of \Spitzer\ and \WISE\ photometry. Alternatively, blending of sources in a particularly unlucky way may result in the non-detection of a real burst.

\section{OUTBURSTS}\label{s.cand}

The only protostars that fulfill all criteria, i.e., HOPS~223 and HOPS~383, have already been discussed in the literature in references such as those given below. Visual inspection confirms that their brightness changes from \Spitzer\ to \WISE\ are reflective of their outbursts, not source confusion in the large \WISE\ beam. These changes are summarized in Table~\ref{t.cand}. HOPS 223 brightened by more than the threshold in all three bands and is marked with a purple circle in Figures~\ref{f.diffmag1} through \ref{f.diff2diff4}. HOPS 383 changed from a non-detection to 14.1 mag at 3.6 \micron, which is consistent with a brightening of at least 2.9 mag given the IRAC sensitivity limit of 17 mag. The minimum change at the shortest wavelength was larger than the 1.0~mag threshold there, and the source also easily satisfied the criteria at the other two wavelengths. It is marked with an orange circle in the relevant figures. Although HOPS 223 and HOPS 383 are both in regions that were mapped by MIPS in 2004 and 2008, the full comparison of \Spitzer\ to \WISE\ data indicates that these are the only two sources that erupted during the 6.5 yr interval. Their fortuitous location at the overlap of the two MIPS fields serves only to confirm the discovery of their bursts.

\begin{deluxetable}{cccccc}
\tablecaption{Outburst Candidates\label{t.cand}}
\tablewidth{\hsize}
\tablehead{\colhead{HOPS} & \colhead{RA} & \colhead{Dec} & \multicolumn{3}{c}{Magnitude change}\\
\colhead{ID} & \colhead{(deg)} & \colhead{(deg)} & \colhead{3.6 \micron} & \colhead{4.5 \micron} & \colhead{24 \micron}}
\startdata
223 & 85.7019 & $-$8.2762 & 1.69 & 2.23 & 1.77 \\
383 & 83.8742 & $-$4.9975 & $>$2.9 & 2.40 & 3.96 \\
\enddata
\end{deluxetable}

\subsection{HOPS 223}\label{s.223}

Our comparison of \Spitzer\ and \WISE\ data was motivated by the large population of protostars in Orion and the announcement by \citet{car11} that HOPS~223, a flat-spectrum protostar in the Lynds 1641 region, had undergone an outburst. The protostar was identified as [CTF93]216-2 in the discovery paper due to its proximity to the protostar [CTF93]216 (\citealt{che93}; HOPS 221 in our survey). It was subsequently given the variable-star identifier V2775 Ori \citep{kaz11}. Here we refer to it by its HOPS number for consistency with the rest of our sample.

\begin{figure}
\includegraphics[width=\hsize]{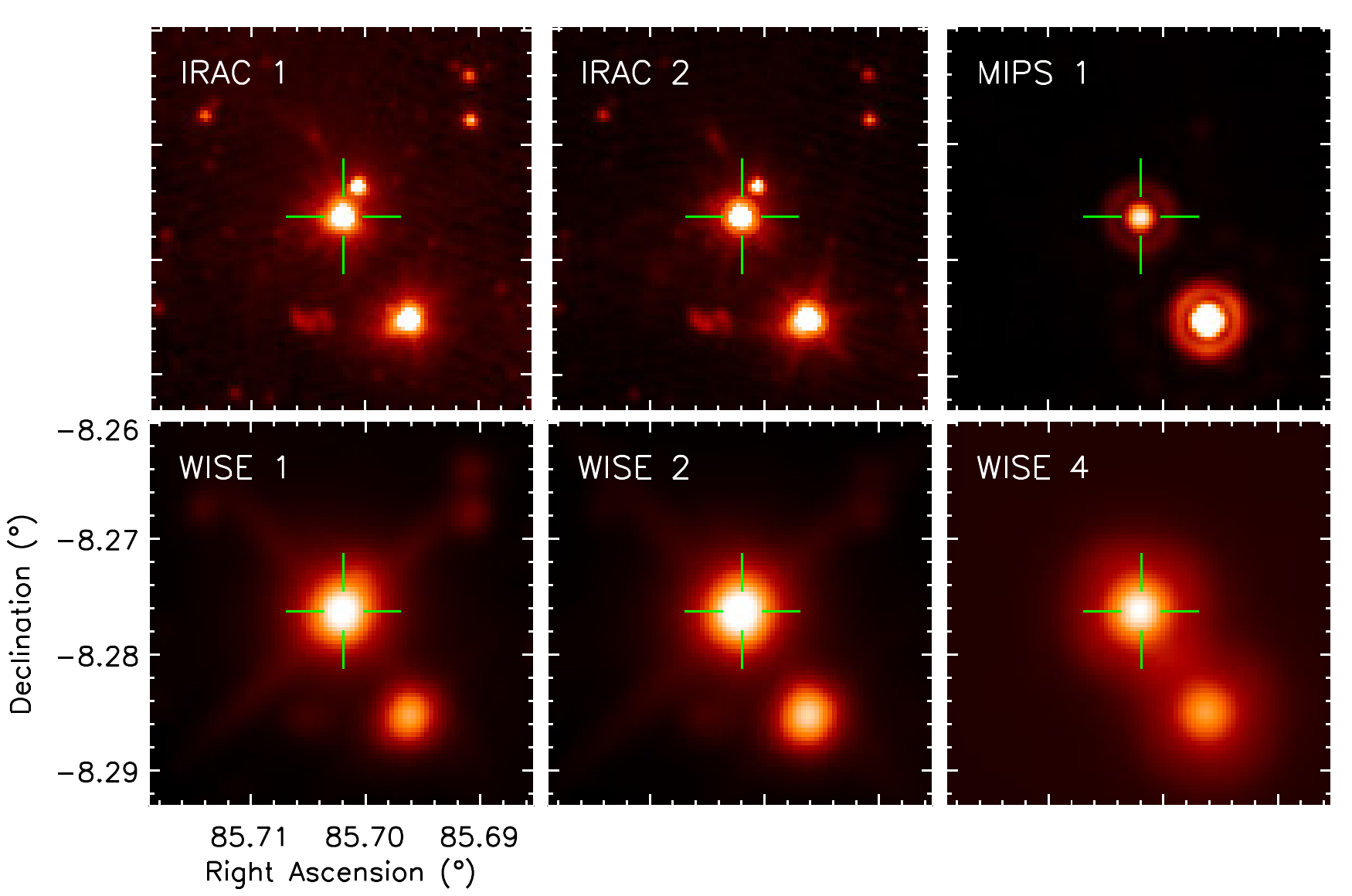}
\caption{HOPS 223 (at the center of the crosshairs) as seen in {\it Spitzer}'s IRAC~1, IRAC~2, and MIPS~1 bands (top row, from left), as well as \WISE's W1, W2, and W4 bands (bottom row, from left).\label{f.223}}
\end{figure}

Our technique recovers the HOPS 223 outburst. The magnitude changes at 3.6, 4.5, and 24 \micron\ were 1.7, 2.2, and 1.8 mag, respectively, corresponding to increases in flux density by factors of 4.7, 7.9, and 5.3. These correspond to magnitude changes of 2.8 to 3.2 standard deviations above the mean. The source offers solid detections in every band at both epochs, and the outburst is obvious in the corresponding images, shown in Figure~\ref{f.223}. Before the outburst, its 3.6 and 4.5 \micron\ flux densities were similar to those of HOPS 221 to the southwest, and at 24 \micron\ it was fainter than HOPS 221. After the outburst, it was the brightest source in the field at all three wavelengths.

Any blending of HOPS 223 with the star 11\arcsec\ to its northwest is insignificant in the post-outburst \WISE\ images. This star was found by \citet{meg12} not to have an IR excess, and it is undetected at 24 \micron. In the pre-outburst \Spitzer\ images, it is increasingly fainter than HOPS 223 with increasing wavelength, going from 2.6 mag fainter at 3.6~\micron\ to $>6.5$ mag fainter at 24~\micron\ if the detection limit there is 9 mag. The contrast would be even greater post-outburst.

Based on photometry and imaging spanning a range of regimes from submillimeter to near-IR, HOPS 223 was determined to have risen in luminosity from 4.5 to $\sim51$~$L_\sun$, a factor of about 12, sometime between 2005 April~2 and 2007 March~12.  \citet{fis12} used criteria developed by \citet{con10} to determine that the near-IR spectrum of HOPS 223 after its outburst was consistent with that of an FU~Orionis object. Though not as extreme as classical examples of FU Orionis outbursts such as FU~Ori itself or V1057 Cyg, the accretion rate of HOPS 223 increased from about $10^{-6}$ to $10^{-5}~M_\sun~{\rm yr}^{-1}$ during the outburst.

\begin{figure}
\includegraphics[width=\hsize]{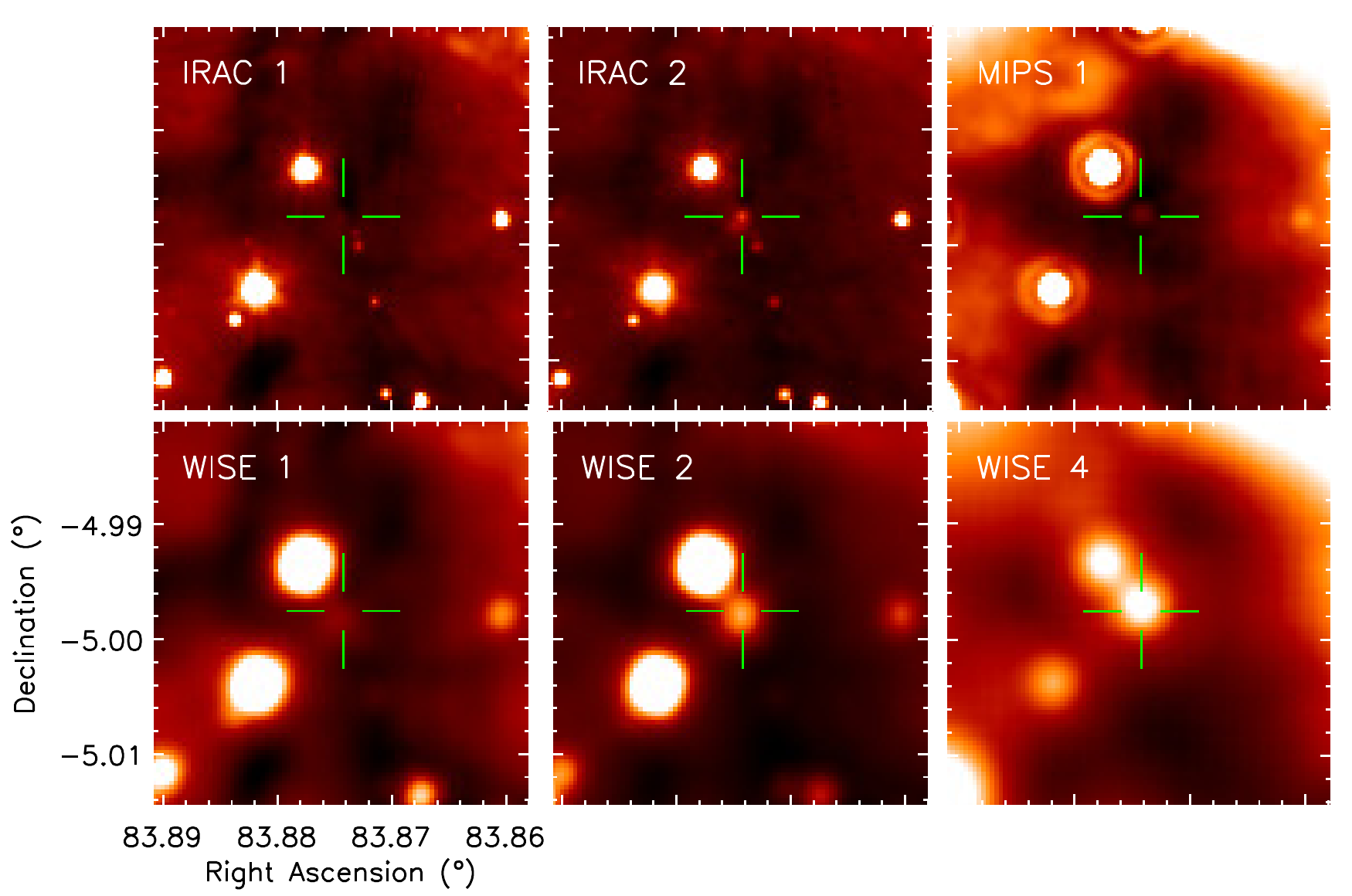}
\caption{HOPS 383 (at the center of the crosshairs) as seen in {\it Spitzer}'s IRAC~1, IRAC~2, and MIPS~1 bands (top row, from left), as well as \WISE's W1, W2, and W4 bands (bottom row, from left).\label{f.hops383}}
\end{figure}

\citet{fis16b} reported additional observations of HOPS 223. In 2015 November they obtained mid-IR photometry with the Stratospheric Observatory for Infrared Astronomy (SOFIA), and in 2016 January they obtained near-IR spectroscopy at Palomar Observatory. These observations, to be discussed in detail in future work, revealed that HOPS 223 had faded slightly but retained the spectral characteristics of an FU Orionis star roughly a decade after its initial outburst.

\subsection{HOPS 383}\label{s.383}

HOPS~383 is a non-detection in the IRAC~1 catalog, but it is firmly detected in all three \WISE\ bands analyzed here. Its magnitude changes at 4.5 and 24~\micron\ were 3.5 and 6.4 standard deviations above the mean, and its magnitude change at 3.6 \micron\ was at least 4.9 standard deviations above the mean. Visual inspection of this source revealed a clear brightening in all three wavelength bands. Images appear in Figure~\ref{f.hops383}.

We previously announced the brightening of this protostar in \citet{saf15}. We estimated that its outburst began between 2004 October 12 and 2006 October 20. Analysis of the HOPS 383 SED, including its detection at submillimeter wavelengths, unambiguously indicates that it is a Class~0 YSO, with 1.5\% of its total luminosity emitted at 350 \micron\ and longer, a bolometric temperature of 43~K, and a peak wavelength of 106~\micron. Its large submillimeter fluxes imply that it is embedded in a massive envelope. Although the coverage of the pre-outburst SED is sparse, the bolometric luminosity was estimated to increase from 0.2 $L_\sun$ to 7.5 $L_\sun$, consistent with a legitimate accretion burst. HOPS~383 appears to be the least evolved source known to undergo an outburst.

Subsequent study of HOPS 383 has revealed a range of behaviors at different wavelengths. \citet{gal15} analyzed archival observations of HOPS 383 at 3 cm. They found two components that are separated by about 0.45\arcsec\ and connected by a ridge of faint emission, likely due to a bipolar jet or a binary. They found only mild variations of the radio flux between 1998 and 2014, interpreting the lack of a radio outburst as evidence that an ejection enhancement did not follow the accretion outburst. \citet{mai17} compared 850~\micron\ data obtained between 2012 August and 2013 August to data obtained at the same wavelength between 2015 December and 2017 Feburary. They found that HOPS 383 was a ``possible'' variable candidate, undergoing a brightness decrease of $2.66\pm0.64\%$ yr$^{-1}$. Imaging of the field at $J$, $H$, and $K_s$ in 2015 December resulted in non-detections \citep{fis17a}, suggesting a decline in the near-IR flux density after the acquisition of post-outburst data between 2008 and 2011. Further study of this deeply embedded outburst is essential for understanding episodic accretion in the youngest protostars.

\subsection{Other Outbursts}

Additional protostellar bursts were known in the Orion molecular clouds before the \Spitzer\ observations that begin our baseline. They are not considered in our statistical treatment, but we review them here for completeness. Just before the beginning of \Spitzer's scientific operations in late 2003, V1647~Ori (HOPS 388) began an outburst that illuminated McNeil's Nebula in Lynds 1630. It peaked in both 2003 and 2008 \citep{rei04,asp11}. This outburst is not cleanly classified as either an FU Orionis or an EX Lupi outburst \citep{muz05,fed07,con17a}, but its detailed behavior is less relevant to the luminosity problem than the finding of an increase in its accretion rate.

Two other Orion protostars brightened well before the epoch of our comparison:\ Reipurth~50 and V883~Ori \citep{str93}. The outburst of Reipurth~50 began between 1955 and 1979, and it has shown recent variability that has been interpreted as variable dust obscuration, not a change in its accretion rate \citep{chi15}. The protostar was saturated in the IRAC and MIPS images used to define the HOPS sample, so it does not have a HOPS number. No precise estimate exists for the beginning of the V883 Ori outburst, but it has been bright since at least 1888 \citep{str93}. It has the number HOPS 376.

\section{THE BURST INTERVAL}\label{s.results}

Our search reveals two accretion bursts between the 2004--2005 \Spitzer\ photometry and the 2010 \WISE\ photometry: HOPS 223 and HOPS 383. Based on this finding, here we estimate the rate of accretion outbursts $r_b$ in a single protostar. This is \begin{equation}\label{e.burstgen}r_b=\frac{n_b}{N_p\Delta t},\end{equation} where $n_b=2$ protostars is the number of bursts found within our time interval $\Delta t$, and $N_p$ is the number of protostars in our sample.

The time interval $\Delta t$ is from the first \Spitzer\ observations to the last \WISE\ observations incorporated in the AllWISE photometry. This varies depending on the location of each protostar within Orion. For 311 of the 319 protostars, the first \Spitzer\ images were obtained in 2004 February or March, and the last relevant \WISE\ images were obtained in 2010 September, giving baselines that range from 6.53 to 6.59 yr. For the remaining eight protostars, all in the Lynds 1622 region of Orion~B, the first \Spitzer\ images were acquired in 2005 October, reducing the interval to 4.90 yr. To an appropriate level of precision, $\Delta t=6.5$~yr for the vast majority of the sample, and we use this value for subsequent calculations.

We carried out the comparison for $N_p=319$~protostars. While 48 of them lack \WISE\ counterparts, we interpret that as evidence against a burst. Even the HOPS 383 outburst, which reached a relatively low post-outburst luminosity of 7.5 $L_\sun$, was detected unambiguously in the \WISE\ images. If we were to use only the 271 protostars with \WISE\ counterparts in the denominator of Equation~\ref{e.burstgen}, we would derive a burst interval that is 15\% shorter, which is a small change compared to the uncertainties discussed below. Inserting the values presented here into Equation~\ref{e.burstgen} gives $r_b=9.65\times10^{-4}$~yr$^{-1}$ and an interval between bursts of $I_b=1/r_b=1036.75$~yr.

To determine the uncertainty in the estimated interval, we model the bursts as a Poisson process over the time interval $\Delta t$ with a rate equal to $1/I_b$. If the burst interval is $I_b$~yr and we watch a protostar for $\Delta t$~yr, the chance we will detect one or more bursts from a given protostar is $1 - e^{-\Delta t/I_b}$. The likelihood $P$ of observing exactly $n_b$ bursts in $N_p$ protostars is then a binomial distribution, where \begin{multline}\label{binomial}P\left(n_b | I_b, N_p, \Delta t\right) = \\ \frac{N_p!}{n_b!(N_p-n_b)!}(1 - e^{-\Delta t/I_b})^{n_b}(e^{-\Delta t/I_b})^{(N_p-n_b)}.\end{multline} Adopting uniform priors, this gives the posterior probability $P\left(I_b | n_b, N_p, \Delta t\right)$ for the burst interval.

\begin{figure}
\includegraphics[width=\hsize]{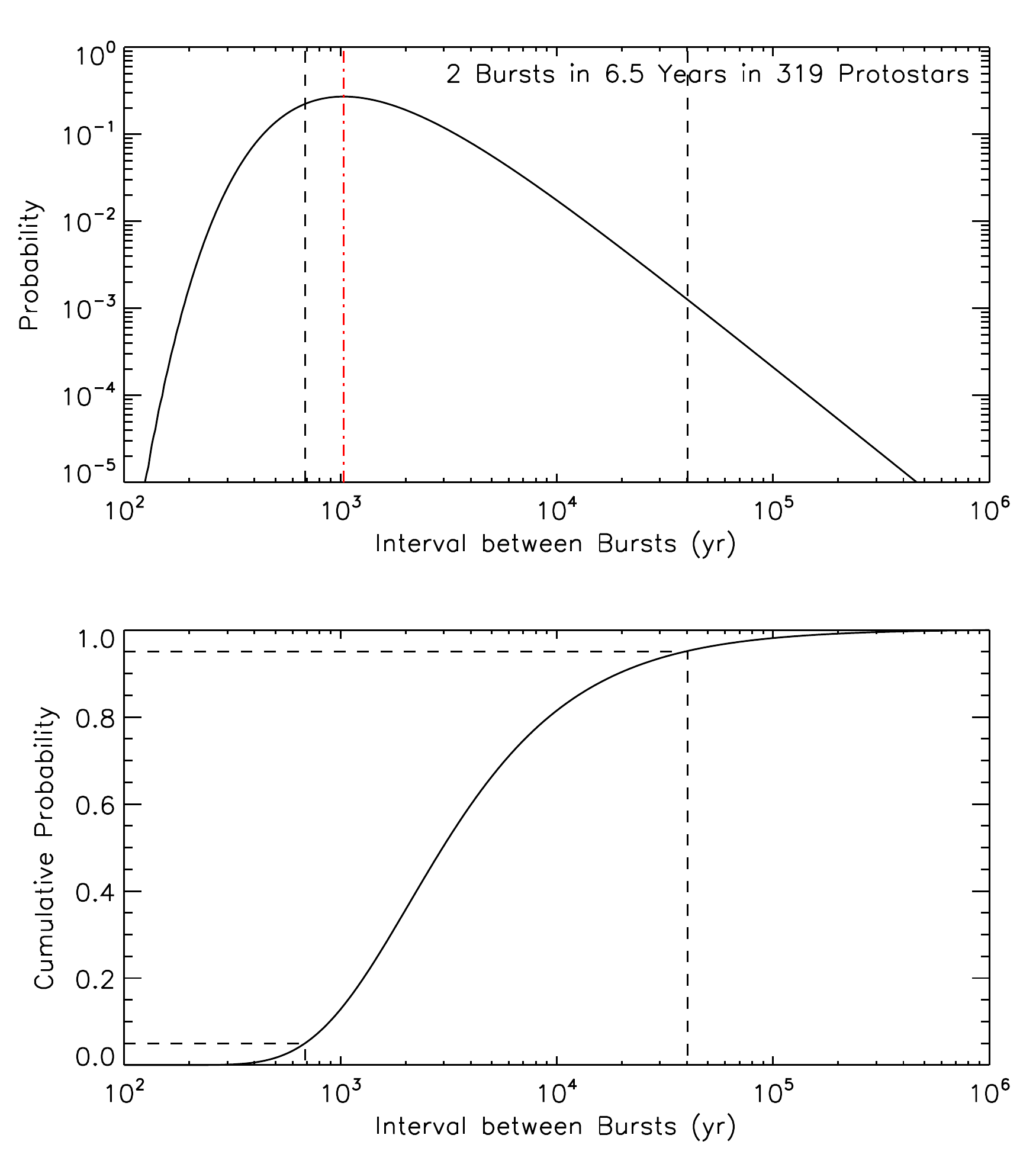}
\caption{{\em Top:} Probability of observing exactly 2 outbursts in 319 protostars over a 6.5~yr window as a function of outburst interval. The probability peaks at about 1040~yr (red dot-dashed line). Dotted lines illustrate the 90\% confidence interval, which is from 690 to 40,300~yr. {\em Bottom:} The cumulative probability distribution, which further illustrates the confidence interval.\label{f.prob}}
\end{figure}

The top panel of Figure~\ref{f.prob} plots $P$ against $I_b$ for our observational result of $n_b=2$ bursts in $N_p=319$ protostars over $\Delta t=6.5$ yr. The probability peaks at 27.2\% for an interval of 1033.50 yr. This evaluation of the probability distribution function for $I_b$ gives a most likely interval that is very slightly less than the interval estimated above; henceforth, we will use 1040 yr. The 90\% confidence interval, further illustrated with the cumulative probability distribution in the bottom panel, is from 690 to 40,300 yr; i.e, there is a 5\% chance that the interval is below this range and a 5\% chance that the interval is above it. For the $N_p=271$ case discussed above, where we exclude the \Spitzer\ protostars without \WISE\ counterparts, the most likely interval is 880 yr with a 90\% confidence interval extending from 586 to 34200 yr, all about 15\% shorter than the adopted values.

The broad 90\% confidence interval is consistent with the findings of \citet{hil15}, who demonstrated that a sample of far more than 319 protostars is needed to precisely estimate the burst interval with high confidence given the available time intervals. If $I_b$ is of order one thousand years and the protostellar lifetime is $5\times10^5$ yr \citep{eva09}, then a given protostar will undergo several hundred outbursts.

\section{DISCUSSION}\label{s.discuss}

\subsection{Dependence of Outburst Interval on Evolution}

The typical burst interval of 1040~yr in Section~\ref{s.results} is much less than the interval calculated from results of similar analyses of Class II YSOs. Comparing \WISE\ and \Spitzer\ data for samples of several thousand primarily Class II YSOs in various star-forming regions, \citet{sch13} used criteria similar to our own to find that the most likely burst interval is between 5000 and 50,000~yr. \citet{car01} analyzed near-IR photometry spread across 2 yr of a region centered on the Orion Nebula Cluster with 2700 YSOs and detected no bursts. This puts a lower limit on the burst interval of 5400~yr. Though the upper end of our 90\% confidence interval does overlap with the lower limits of these studies, the difference is intriguing.

These findings are consistent with outburst modeling where the disk instabilities are ultimately due to mass infall from the protostellar envelope. In these models most bursts would be expected to occur in Class 0 and Class I protostars, where the remaining envelope mass is significant, and bursts would be possible but rare for Class II YSOs, where the envelope is at most tenuous. Consistent with this outline, \citet{vor15} count bursts that occur in their gravitational instability simulations and find that most bursts occur in Class I protostars, while a smaller fraction take place in Class~0 protostars, and only a few happen in Class II YSOs.

\subsection{Protostellar Outbursts and the Luminosity Problem}

To consider the implications of our findings for the luminosity problem, we use a toy model to explore the fraction of a star's ultimate main-sequence mass that is accumulated in outbursts. If the protostellar phase can be divided unambiguously into periods of quiescence and outburst, and the fraction of the time spent in outburst is $f_b$, then the total mass accumulated in quiescence is $M_q=\left(1-f_b\right)\int \dot{M_q}\,dt$. If the accretion rate during bursts is a constant factor $A$ greater than the slowly varying accretion rate during quiescence, then the mass accumulated during outbursts is $M_b=Af_b\int \dot{M_q}\,dt$. In both cases the integral is over the duration of the protostellar phase.

\begin{figure}
\includegraphics[width=\hsize]{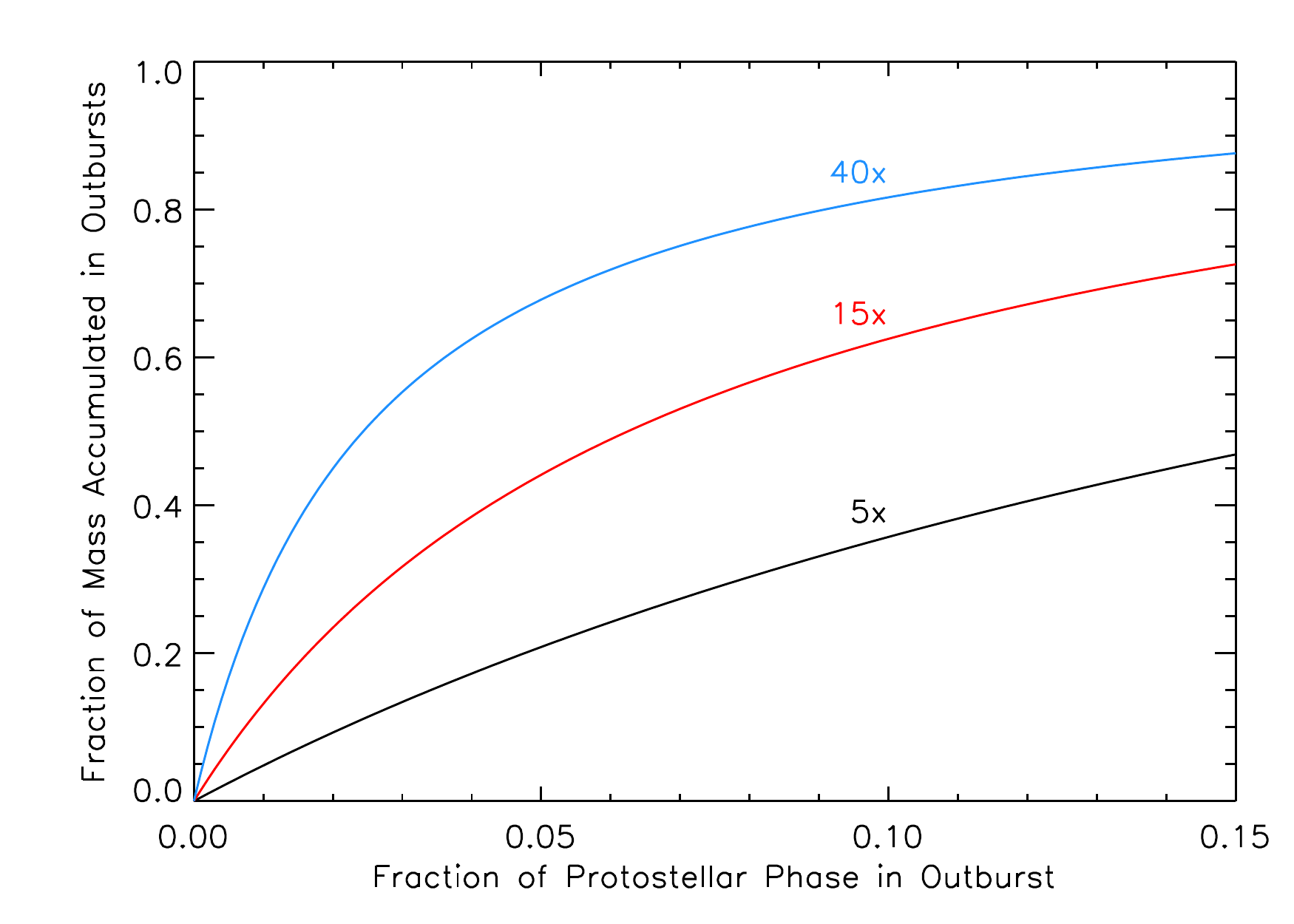}
\caption{Fraction of the stellar mass accumulated during outbursts as a function of the fraction of the protostellar phase spent in outbursts. Each curve corresponds to a case where the accretion rate during outbursts exceeds that of the accretion rate in adjacent periods of quiescence by the indicated factor. The indicated ratio of post-to-pre-burst accretion rates is an upper bound to the ratio of post-to-pre-burst luminosities. \label{f.mass}}
\end{figure}

The fraction of the mass accumulated in bursts is then $M_b / \left(M_b + M_q\right)$, which is simply $Af_b/\left[\left(A-1\right)f_b+1\right]$. This quantity is plotted against $f_b$ for three values of $A$ in Figure~\ref{f.mass}. The fraction $f_b$ is $d_b/I_b$, the duration of each outburst divided by the interval between their onsets. We have argued that $I_b$ is of order 1000 years, but the $d_b$ of the outbursts we have examined are as yet unknown, since they are still ongoing. SOFIA observations of the HOPS 223 and 383 outbursts indicate that they were still at elevated luminosities about a decade after their onsets (W.~J. Fischer et al., in preparation). 

The outbursts included in our calculation of the burst frequency experienced luminosity increases of factors 12 (HOPS 223) and 35 (HOPS 383). These are lower limits to the factors by which the mass accretion rates increased, although we argued in Section~\ref{s.crit} that the two factors are similar if the accretion luminosities are much greater than the stellar luminosities, as expected for protostars. If outbursts like that of HOPS 223 are typical, representing fifteen-fold increases in the mass accretion rate and persisting for 20 to 40 years ($0.02<f_b<0.04$), then between 23\% and 38\% of a star's mass would be accumulated in outbursts. These values are larger than but not dramatically different from the fraction of 25\% estimated by \citet{off11}. If the HOPS 383 outburst is typical, then forty-fold increases in the mass accretion rate are more likely, and more than half of a star's mass would be accumulated in outbursts for any $f_b>0.024$.

To investigate the luminosity problem, \citet{fis17b} plotted bolometric luminosities and temperatures for 315 Orion protostars. They found that the spread in luminosities at each bolometric temperature is three orders of magnitude, and that this spread is consistent with a slowly, exponentially decreasing accretion rate that depends on the final mass. \citet{off11} also found that models with an approximately constant accretion time and a limited role for outbursts could produce a broad spread in luminosities. Even if it is responsible for the accumulation of a nontrivial percentage of a star's main-sequence mass, episodic accretion is not required to explain the spread in BLT diagrams or to solve the luminosity problem, which assumes a constant, mass-independent accretion rate.

Finally, if outbursts that result in luminosity increases of factors of 10 to 100 do occur with the frequency we calculate here, then these excursions are small compared to the factor of 1000 spread in protostellar luminosities. They do, however, imply that the path of each protostar across the BLT diagram does not vary smoothly with age and mass, and that stars with the same main-sequence mass may not have followed the same trajectory across the BLT diagram.

\subsection{Other Constraints on the Protostellar Outburst Frequency}

The separation of knots in protostellar outflows is an alternative way to estimate the frequency of enhanced accretion events. \citet{arc13} observed clumps in the HH~46/47 molecular outflow, which is driven by a Class~I protostar. They inferred from the spacing and velocities of three clumps that they were ejected approximately 300 yr apart. If these are due to episodic ejection events that are driven by accretion, this points to an outburst interval in this protostar that is at the low end of the range we derive for the Orion protostars.

Another way to estimate the outburst frequency is to look for evidence of past outbursts in freeze-out chemistry \citep{vis15}. Enhanced luminosity alters the chemistry of the protostellar envelope, evaporating CO throughout the envelope and H$_2$O to a radius 10$\times$ larger than usual. The CO evaporation enhances the abundance of HCO$^+$ and reduces the abundance of N$_2$H$^+$; this altered chemistry persists for 10$^3$ to 10$^4$ yr. \citet{jor13} used such arguments to conclude that the protostar IRAS 15398$-$3359 underwent an accretion outburst 100 to 1000 yr ago. The composition of circumstellar dust may also be altered by outbursts. \citet{pot11} detected crystalline silicates in the spectrum of HOPS~68, the first such detection in a cold protostellar envelope. They argued that the heat necessary to crystallize the silicates must have been provided by an episodic accretion event.

A significant reduction in the uncertainty of the interval requires a timescale or a sample that is at least an order of magnitude larger \citep{hil15}, including protostars at distances of up to $\sim2$~kpc instead of only 500 pc. Progress on this is being enabled by near-IR all-sky monitoring from the ground, e.g., with the Visible and Infrared Survey Telescope for Astronomy (VISTA) surveys \citep{con17b} and a United Kingdom Infrared Telescope Infrared Deep Sky Survey (UKIDSS) search for eruptive YSOs \citep{luc17}. It will continue with the planned Wide-Field Infrared Survey Telescope (WFIRST).

While surveys at optical and near-IR wavelengths are valuable, mid-IR to submillimeter surveys provide more direct evidence of changes in the luminosities of the most deeply embedded protostars. Submillimeter monitoring with the James Clerk Maxwell Telescope is proving fruitful in understanding the occurrence rate and duration of outbursts \citep{her17,joh18}. A sensitive far-IR space telescope with fast mapping capabilities such as the proposed {\em Origins Space Telescope} \citep{mei17} would probe deeply embedded protostars at the peaks of their SEDs. \citet{bil12} showed the value of far-IR monitoring in the vicinity of the Orion Nebula, where, for an initial look at the far-IR variability of protostars, they monitored 43 protostars with \Herschel\ over six weeks to find that variations at the 10--20\% level were typical at 70 and 160 \micron. SOFIA is also essential for mid-to-far-IR follow-up of protostellar outbursts to determine their durations and to extend their SEDs across the mid- and far-IR regimes for accurate luminosities.

\section{CONCLUSIONS}\label{s.conc}

We compared three-band \Spitzer\ and \WISE\ photometry for 319 Orion protostars to search for luminosity outbursts in the $\sim6.5$ yr between the acquisition of the two data sets. Although catalog photometry often indicates minor brightening in one or two bands, visual inspection reveals that these are usually false positives due to confusion in \WISE. Those protostars that brightened by at least two standard deviations more than the mean difference at all three wavelengths of comparison were identified as outbursts and visually confirmed as such.

This approach turned up one previously known outburst, V2775~Ori (HOPS 223), and one new burst, HOPS~383, which was announced in our earlier paper \citep{saf15}. No other new sources were discovered. The latter is the first Class~0 accretion burst to be discovered. We estimate the most likely interval between bursts in a given protostar to be 1040~yr, with a 90\% chance that the interval lies between 690 and 40,300~yr.

This and related studies of more evolved, Class II YSOs paint a picture where outbursts are frequent in the protostellar stage, occuring once every $\sim10^3$ yr, and they are much less frequent in the optically revealed, disk dominated stage, occuring every $\sim10^4$ yr. A similar dependence on evolutionary class was also reported by \citet{bar05} and \citet{con17b}. There is emerging evidence that protostellar outbursts are less dramatic than Class II outbursts; those reported here represent brightenings of factors of 10--40 rather than the best known FU Orionis events observed in Class II YSOs, which feature brightenings by factors of 100 or more. These details are consistent with the predictions of models where the disk becomes unstable due to infall from a surrounding envelope, it fragments, and the accretion rate spikes when these fragments are accreted onto the star.

The duration of the bursts is not well constrained, but it appears to be typically longer than a decade. For a range of assumptions, episodic accretion events account for 25\% or more of a star's ultimate main-sequence mass, but they are not required to explain the luminosity distributions of large samples of protostars. Further surveys using a variety of observational techniques are needed to reduce the large uncertainty in the outburst interval.

\acknowledgments

The National Aeronautics and Space Administration (NASA) provided support for this work through awards issued by the Jet Propulsion Laboratory/California Institute of Technology (JPL/Caltech). Support for the contributions of WJF was provided in part by NASA through awards 03\_0143 and 04\_0134 issued by the Universities Space Research Association. This work is based in part on observations made with the \textit{Spitzer Space Telescope}, which is operated by JPL/Caltech under a contract with NASA. It also makes use of data products from the \textit{Wide-field Infrared Survey Explorer}, which is a joint project of the University of California, Los Angeles, and JPL/Caltech, funded by NASA. Finally, it has made extensive use of the NASA/IPAC Infrared Science Archive, which is operated by JPL/Caltech under a contract with NASA.

\end{document}